\documentclass[5p,times,twocolumn]{elsarticle}

\usepackage{amsmath,amssymb}
\usepackage{graphicx}
\usepackage{booktabs}
\usepackage{siunitx}
\usepackage{bm}
\usepackage[T1]{fontenc}
\usepackage[utf8]{inputenc}
\usepackage{microtype}
\usepackage{lineno}
\usepackage{url}
\usepackage{tikz}
\usetikzlibrary{positioning, arrows.meta}
\journal{Acta Astronautica}

\begin{document}

\begin{frontmatter}

\title{A simplified engineering algorithm for collision risk assessment and classification of LEO satellites}

\author[cisas]{Federico Toson\corref{cor1}}
\ead{federico.toson@unipd.it}

\cortext[cor1]{Corresponding author.}
\affiliation[cisas]{organization={CISAS ``G. Colombo'', University of Padova},
  addressline={via Venezia 15}, postcode={35131}, city={Padova}, country={Italy}}

\begin{abstract}
Collision risk in LEO is now a sustainability constraint, not only a mission-level concern. Current risk-assessment standards (ECSS, NASA) classify risk in discrete categories per single mission and produce no continuous index aggregable across missions or feedable into broader sustainability assessments. This paper develops an engineering algorithm in which the expected number of collisions is estimated from the kinetic theory of the debris flux and combined with a severity term to give a continuous, dimensionally consistent risk index. An individual index, weighted by replacement cost, captures the expected loss to the operator; a collective index with homogeneous per-target severity captures the burden imposed on the operational population at the mission altitude band, independent of the mission's own value. Applied with a May~2026 Celestrak catalogue snapshot to three representative cases (ENVISAT, Sentinel-6, Starlink V2~Mini), the collective index exceeds the individual one by over two orders of magnitude for the constellation element: operator-centred assessment understates the collective cost of large constellations. A random sample of $100$ catalogued LEO satellites, swept over $100$ random seeds, confirms the asymmetry as structural across the constellation and Earth-observation tiers of the catalogue. The construction is checked internally against a snapshot proxy of the ECOB family (rank correlation $\rho_{S}=0.91$ on a joint sample of $103$ objects) and externally against published ECOB values for MetOp-A and Sentinel-2, which the proxy matches within a factor of two; the asymmetry survives the swap to a mass-based severity weight, which widens, not closes, the constellation gap. A parametric analysis then gives an explicit replacement-cost threshold per altitude band, approximately one to five million USD across the operational LEO bands, below which any satellite is structurally underestimated by operator-centred frameworks. The index is released as an open MATLAB toolchain and is intended as the orbital component of a multi-domain sustainability assessment.
\end{abstract}


\end{frontmatter}

\section*{Nomenclature}
\noindent\footnotesize
\begin{tabular}{@{}lp{0.70\linewidth}@{}}
$\alpha$ & fragmentation-weight exponent ($=0.5$) \\
$A_m$, $A_{ref}$ & total surface of $m$, reference \\
$C_m$, $C_{ref}$ & replacement cost of $m$, reference \\
$C^{*}(h)$ & threshold for asymmetry $\geq\!10^{2}$ \\
$f_{frag}$ & fragmentation-severity weight \\
$h_m$ & mission perigee altitude \\
$K(h)$ & intercept of asymmetry--cost relation \\
$N(h_m,\Delta h)$ & catalogued objects in altitude shell \\
$P_m^{ind}$, $P_m^{col}$ & individual, collective impact counts \\
$R_m^{ind}$, $R_m^{col}$ & raw individual, collective risk \\
$\tilde{R}_m^{\,s}$ & normalised risk, $s\in\{ind,col\}$ \\
$S_m^{ind}$, $S_m^{col}$ & individual, collective severity \\
$s_0$ & homogeneous per-target harm (cancels in normalisation) \\
$T_{op}$, $T_{res}$, $T_{tot}$ & operational, residual, total lifetimes \\
$V_{shell}$ & altitude-shell volume \\
$v_{rel}$ & mean relative LEO velocity \\
$\Delta h$ & altitude-shell half-width \\
$\rho_{all}$, $\rho_{op}$ & catalogued, operational densities \\
$\sigma_m$ & collision cross-section (frontal area) \\
\end{tabular}
\normalsize
\vspace{0.8em}

\section{Introduction}

The Low Earth Orbit (LEO) environment has changed sharply over the past decade. Catalogued objects above ten centimetres now exceed thirty thousand, the population between one and ten centimetres is estimated above one million, and the active satellite count has more than tripled in the past five years; projections place over one hundred thousand spacecraft in LEO by the early 2030s \citep{ESA_SpaceEnv_2024,NASA_ODPO_2024}. Large-constellation deployments \citep{OlivieriFrancesconi2020}, lower launch costs and a wider base of state and commercial operators have moved LEO from a lightly used scientific commons toward an industrial infrastructure.

In this regime, collision risk has shifted from a marginal concern to a long-term sustainability constraint. The cascade dynamics described by Kessler and Cour-Palais~\citep{KesslerCourPalais1978} are no longer hypothetical: the 2007 Fengyun-1C anti-satellite test, the 2009 Iridium-33/Cosmos-2251 collision and the 2021 Cosmos-1408 event injected the four fragment clouds that still dominate the tracked debris at their altitudes (see Section~\ref{sec:cases}). Conjunction events, avoidance manoeuvres and the rising premiums of in-orbit insurance all indicate that the most contested altitude bands are approaching, in some scenarios, the conditions for self-sustained debris growth \citep{LiouJohnson2006,Krag2017,ESA_SpaceEnv_2024}. The same trend has motivated dedicated monitoring technologies, including sub-millimetre debris detectors on student platforms \citep{TosonJ2050}.

Current practice is predominantly \emph{categorical}. The frameworks formalised in the standard of the European Cooperation for Space Standardization (ECSS)~\citep{ECSS_M_ST_80C} and in the corresponding NASA documentation~\citep{NASA_STD_8719} combine likelihood and severity into discrete risk classes arranged in a colour-coded matrix. These frameworks suit the qualitative communication of risk but present three limitations as the contested regions saturate: the discrete classification compresses order-of-magnitude variation into a few classes; it does not aggregate across missions; and it does not produce a continuous index that can feed broader sustainability frameworks such as the Life Cycle Assessment methodology under development at ESA \citep{PEFCR4Space2025} or the orbital indices led by the Environmental Consequences of Orbital Breakups (ECOB) family \citep{Letizia2016}.

This paper proposes a continuous risk index that bridges categorical engineering practice with continuous sustainability indices, and uses it to identify an explicitly quantified regulatory blind spot. The construction rests on a single methodological choice: a \emph{homogeneous collective severity}, independent of the mission's own replacement value, which differs from the asset-weighted aggregation customary in collective risk accounting and is what makes the operator-versus-environment asymmetry quantitatively visible. The novelty lies in this framing and in the closed-form regulatory consequences derived from it, not in the underlying kinetic-flux probability model, which is classical~\citep{KesslerCourPalais1978}. Built on this framing and on earlier work by the author \citep{Toson2022}, the paper contributes: the homogeneous-severity collective index and its classification thresholds, aggregable across missions and benchmarked against a snapshot proxy of the ECOB family at Spearman $\rho=0.91$ on a joint sample of $103$ catalogued LEO objects and against published ECOB values for MetOp-A and Sentinel-2 to within a factor of two; a statistical evaluation on a random sample of one hundred satellites showing the asymmetry holds for every sampled object and clusters in the constellation altitude band, followed by a parametric analysis giving an explicit replacement-cost threshold per altitude band below which current debris-mitigation guidelines structurally underrate the mission; and an open MATLAB toolchain with a 2026 Celestrak calibration. The index is designed as the orbital component of a multi-domain Space Sustainability Composite Indicator: this paper is the first contribution of a wider research effort by the author on multi-domain sustainability of space systems, of which the atmospheric-impact and terrestrial life-cycle components are in preparation.

\section{Background and state of the art}

\subsection{Probabilistic foundations of conjunction analysis}
The quantitative assessment of collision risk is grounded in the probabilistic treatment of conjunction events, expressing the probability of collision during a close approach as the integral of a bivariate Gaussian over the projected hard-body region \citep{FosterEstes1992,Alfano2005,Chan2008,Klinkrad2006}. For short-term LEO encounters the formulation reduces to a closed form \citep{AkellaAlfriend2000}, and Monte Carlo treatments provide validation benchmarks \citep{LemmensKrag2014}. These methods produce a numerical probability for an individual encounter; their mapping into a risk index is performed downstream and is not standardised.

\subsection{Categorical risk classification}
The dominant practice combines collision probability with a severity score into a categorical magnitude classification \citep{ECSS_M_ST_80C,NASA_STD_8719}. The categorical structure compresses substantial variation into a small number of classes, is designed for individual missions and does not aggregate across them, and does not produce a continuous index integrable within sustainability frameworks.

\subsection{Continuous orbital sustainability indices}
In parallel, the orbital sustainability community has developed continuous metrics quantifying a mission's contribution to the orbital environment, and they span a fidelity ladder. At the high-fidelity end, the ECOB index \citep{Letizia2016,Letizia2017}, subsequently extended to fragmentation severity and disposal scenarios \citep{Letizia2019,Letizia2023}, evaluates the environmental consequence of a breakup through long-term debris-evolution modelling; a recent comparative assessment maps the object-based index landscape \citep{Letizia2026comp}. Below it sits a family of deliberately simple formulations, each weighing a different proxy of harm: the pre-fragmentation collision mass of the Criticality of Spacecraft Index \citep{Rossi2015}, whose line has since produced a fragmentation environmental index resolving individual breakup events \citep{Gisolfi2025}; expected fragment counts, the proxy adopted here; and the residence-time family of object-years and object-kilogram-years, with density-weighted permutations. Capacity-based constructions rank collision risk against the carrying capacity of the orbital shells and discretise it into rating bins \citep{Muciaccia2026}, screening studies rank the statistically most concerning derelicts directly \citep{McKnight2021} and derive closed-form space-domain-awareness estimates \citep{McKnight2023}, and stability analyses estimate the critical number of spacecraft a shell sustains before debris growth becomes self-sustained \citep{LewisKessler2025}. The MITRI risk index ranks active-debris-removal targets through long-term Monte Carlo evolution \citep{Servadio2024,Simha2025}, environment-capacity ratings support space traffic management \citep{Lemmens2020}, and, at the institutional level, the Space Sustainability Rating composite indicator \citep{Rathnasabapathy2025} aggregates several of these into a single mission rating. These indices quantify a single mission's contribution to a collective long-term outcome; what none of them carries is the operator's own expected loss alongside it.

\subsection{The gap and the contribution of this work}
The categorical engineering tradition and the continuous sustainability tradition are individually mature but do not bridge: the translation between them is, in current practice, ad hoc. The contribution of this paper is a specific pairing across that divide, not the discretisation of a continuous metric, which is established practice \citep{Muciaccia2026}, and not the severity proxies themselves, which are deliberately conventional. The algorithm carries two normalised indices through the same kinetic-flux backbone: the risk a mission bears, cost-weighted as in operator-centred engineering practice, and the risk it imposes on the operational population of its band, with the homogeneous severity of an environment-centred view. Either axis alone exists in the literature reviewed above; evaluating both on the same population, normalisation and snapshot is what makes their divergence measurable, and Section~\ref{sec:cases} shows that divergence exceeds two orders of magnitude exactly where the regulatory stakes are highest. Around this core, the construction retains classification thresholds compatible with the ECSS five-class scheme and a continuous index aggregable across missions. A complementary open-data screening contribution \citep{Poggi2026} classifies \emph{deorbiting compliance} via machine learning on fused public catalogues; the present construction is distinct, quantifying \emph{collision risk} via a physics-based kinetic-flux formulation. Both are complementary to high-fidelity tools such as the ECOB index \citep{Letizia2016} and long-term debris-evolution models \citep{LiouJohnson2008}, to which borderline cases can be routed. The closest risk-index work \citep{Servadio2024,Simha2025} ranks debris populations from the ADR perspective via long-term Monte Carlo evolution; the present construction is mission-side and prospective, with a closed-form classification per altitude band evaluated on a catalogue snapshot.

\section{Methodology}

\subsection{Conceptual model and definitions}
Two complementary scenarios are considered. The \emph{individual} scenario assesses the collision risk borne by a specific mission $m$ over its operational lifetime, $R_m^{ind}$; the \emph{collective} scenario assesses its contribution to the cumulative risk imposed on the operational population of its altitude band, $R_m^{col}$. Both share the structure of classical engineering risk assessment,
\begin{equation}
R = P \cdot S,
\end{equation}
where $P$ is the expected frequency of the adverse event and $S$ the severity of its consequence. The scenarios differ in \emph{whose} risk is measured: the individual index quantifies the expected loss to the operator, while the collective index quantifies the environmental burden imposed on others. This distinction is the source of the asymmetry revealed in Section~\ref{sec:cases}.

\subsection{Kinetic-flux probability}
\label{sec:flux}
Instead of computing discrete geometric conjunctions, which are numerically fragile in sparsely populated bands, the probability is estimated through the kinetic theory of the debris flux, the formulation underlying both classical debris-environment models \citep{KesslerCourPalais1978} and the ECOB index. The instantaneous collision rate experienced by an object of cross-section $\sigma$ moving with mean relative velocity $v_{rel}$ through a region of spatial number density $\rho$ is
\begin{equation}
\dot{n} = \rho \cdot \sigma \cdot v_{rel}.
\end{equation}
The spatial density in the altitude shell of half-width $\Delta h$ centred on the mission perigee altitude $h_m$ is computed directly from the catalogued population,
\begin{equation}
\rho(h_m) = \frac{N(h_m,\Delta h)}{V_{shell}(h_m)},
\label{eq:density}
\end{equation}
with $V_{shell}(h_m) = \tfrac{4}{3}\pi\!\left[(R_\oplus + h_m + \Delta h)^3 - (R_\oplus + h_m - \Delta h)^3\right]$,
where $N(h_m,\Delta h)$ is the number of catalogued objects with perigee in the shell and $R_\oplus$ the Earth radius. The mission collision cross-section $\sigma_m$ is taken as its exposed (frontal) projected area, the convention of ESA's Debris Risk Assessment and Mitigation Analysis (DRAMA) tool; for a tumbling convex body the omnidirectional mean is $A_{total}/4$ by Cauchy's formula, so absolute fluxes are upper bounds while the normalised ratios of Section~\ref{sec:riskindex} are insensitive, the reference sharing the same convention. Throughout, the mean relative velocity is fixed at $v_{rel}=10\,\mathrm{km\,s^{-1}}$, the conventional LEO value consistent with the $9.65 \pm 0.88\,\mathrm{km\,s^{-1}}$ estimate of Rossi and Farinella~\citep{RossiFarinella1992}. Centring the shell on the perigee altitude requires no propagation and reads directly off the catalogue; for the near-circular orbits that dominate the active LEO population it coincides with the residence altitude, and for eccentric objects it concentrates the object at the densest crossing of its orbit, a simplification consistent with the screening purpose (a dwell-time-weighted average across shells is the natural refinement; for near-circular orbits the two coincide). The half-width $\Delta h = 25\,\mathrm{km}$ is half of the 50~km bin standard adopted in density-based debris assessments~\citep{Letizia2016}: wide enough that the shell populations at the altitudes of interest count hundreds to thousands of objects, so the density estimate does not ride on small counts, and narrow enough to resolve the vertical structure of the Starlink and sun-synchronous shells visible in Figure~\ref{fig:altitude}. The population entering~\eqref{eq:density} is the current catalogue: the index is a snapshot instrument by design, recomputable as the environment evolves, and makes no claim about projected future traffic, which is the domain of the long-term evolution models it is benchmarked against in Section~\ref{sec:ecob-bench}. The \emph{individual} probability is the expected number of impacts on $m$ during its operational life $T_{op}$,
\begin{equation}
P_m^{ind} = \rho_{all}(h_m)\,\sigma_m\,v_{rel}\,T_{op},
\end{equation}
with $\rho_{all}$ the shell density~\eqref{eq:density} evaluated over \emph{all} catalogued objects, operational and debris alike. The \emph{collective} probability is the contribution of $m$ to the risk of the operational population,
\begin{equation}
P_m^{col} = \rho_{op}(h_m)\,\sigma_m\,v_{rel}\,T_{tot},
\end{equation}
where $\rho_{op}$ is the same density restricted to operational objects and $T_{tot}=T_{op}+T_{res}$ adds the residual orbital lifetime, taken as an input from the mission's disposal declaration or a decay analysis; the ballistic coefficient of the object enters the construction only through that input. $P_m^{col}$ is thus a pure expected count of collisions with the operational population. How much debris such a collision generates, and therefore how much of the surrounding population it threatens, is carried by the fragmentation weight $f_{frag}$, which enters the collective construction through the severity term of Section~\ref{sec:severity}: a larger object produces a larger fragment cloud. It is taken as $f_{frag}=(A_m/A_{ref})^{1/2}$, with $A_m$ the total surface of $m$ and $A_{ref}$ that of the reference mission. Area is preferred to the customary mass proxy because it is observable when mass is proprietary, and because for the similar-density bodies that dominate the catalogue the two track the same breakup scaling; the mass-based variant is implemented in the released toolchain, and Section~\ref{sec:ecob-bench} quantifies the effect of the swap. The half-power exponent is conservative relative to the NASA Standard Breakup Model~\citep{Johnson2001}, where the cumulative fragment count scales with target mass as $M^{0.75}$ (equivalent to $A^{1.1}$ for similar-density bodies); the present $\alpha=0.5$ therefore under-weights large objects relative to a breakup-physics-consistent scaling, a conservative bias against the very asymmetry the index is designed to expose. Below the reference size the same exponent errs in the opposite direction and over-weights small-object fragment production; the small-object regime is treated explicitly in the corollary of Section~\ref{sec:parametric}, and the sweep $\alpha\in[0.3,0.75]$ of Section~\ref{sec:limitations} leaves the five-class assignment of every case study unchanged. The formulation is dimensionally consistent: strictly, $P_m^{ind}$ and $P_m^{col}$ are expected impact counts (Poisson means), not probabilities. For a Poisson process the collision probability is $1-e^{-P}$, which coincides with $P$ in the $P\ll 1$ regime typical of individual missions; the symbol $P$ is retained for continuity with the $R = P\cdot S$ structure of the engineering risk standards.

This is a simplified first-order proxy. It does not replace, and does not compete with, the full ECOB index \citep{Letizia2016,Letizia2017}, which remains among the rigorous references for orbital sustainability assessment and relies on detailed long-term debris-evolution modelling. The division of labour follows the inputs and the cost: the present formulation needs a catalogue snapshot and per-mission engineering parameters, runs in seconds, and suits licensing-stage screening, fleet-wide sweeps and the classification of Section~\ref{sec:riskindex}; ECOB-class tools need long-term evolution modelling and detailed mission data, and borderline or high-stakes cases identified by the screening are routed to them. Where a rigorous orbital assessment is required, as in a forthcoming multi-domain framework by the author, the full ECOB index should be used in place of this proxy. The formulation likewise abstracts from collision-avoidance operations; Section~\ref{sec:limitations} explains why the structural load, and not the mitigated residual, is the quantity the regulatory reading prices.

A related boundary is the population the densities count. Both are built from the tracked catalogue, while most mission-ending risk to an operational satellite comes from lethal non-trackable fragments in the one-to-ten-centimetre range, whose LEO population exceeds the tracked one by roughly a factor of thirty \citep{ESA_SpaceEnv_2024}. The two indices are exposed asymmetrically. The collective index is unaffected: its target population, the operational satellites of the band, is tracked by construction. The individual index is a lower bound: extending $\rho_{all}$ to the sub-catalogue population would multiply the individual flux by a lethal-to-tracked ratio $\lambda(h)$. In the normalised index a uniform $\lambda$ cancels exactly, numerator and reference scaling alike, so only the altitude profile of $\lambda$ enters the asymmetry of Section~\ref{sec:cases}, and that profile follows the same debris concentrations that shape the tracked population. The tracked-only choice can therefore attenuate the asymmetry, not produce it, and Section~\ref{sec:limitations} restates the consequence: absolute individual indices are lower bounds, while the classification, evaluated against the same reference population, is preserved.

\subsection{Severity model}
\label{sec:severity}
The severity differs by scenario, and this is the central choice of the construction. For the \emph{individual} scenario, severity is the economic consequence to the operator of losing the mission, operationalised as the replacement value, $S_m^{ind}=C_m$. Replacement value is not the operator's full economic loss: the loss can add foregone revenue during re-procurement, reputational damage and, for an ageing bus, a replacement dearer than the residual service it restores \citep{Adilov2015,Rao2020}. It is adopted here because it is the component that is observable, comparable across missions and insurable, and because the choice is least distorting exactly where the asymmetry of Section~\ref{sec:cases} matters: for elements of a redundant constellation the marginal service loss of one unit is small by design, so replacement value approaches the full marginal loss, while for monolithic missions it understates the loss and the individual index reads as a lower bound. In a regulatory setting $C_m$ would be the declared replacement value, anchored to the insured value where the asset is insured and to the unit costs disclosed in licensing and financial filings otherwise, since large constellations often self-insure; the incentive properties of the declared figure are examined in Section~\ref{sec:discussion}. A high impact probability on a low-value object thus yields only a moderate expected loss to the operator. For the \emph{collective} scenario, severity factors into two parts, $S_m^{col}=s_0\,f_{frag}$: a per-target harm $s_0$, taken \emph{homogeneous} across missions because the loss the commons suffers per struck satellite does not depend on the value of $m$, and the fragment-production weight $f_{frag}$ of Section~\ref{sec:flux}, which carries how much debris each collision seeds. Each factor keeps its physical role: the probability term counts collisions with the operational population, the severity term weighs their consequence. This choice differs from the asset-weighted aggregation customary in financial risk accounting: it deliberately decouples the environmental burden from the operator-level asset value, and is the design point that makes the divergence between operator-centred and environment-centred assessment explicit and quantifiable. The constant per-target value cancels in the normalisation, so a cheap object can carry a very high collective risk when it operates in a congested band. Section~\ref{sec:parametric} characterises this divergence parametrically.

\subsection{Risk index and classification thresholds}
\label{sec:riskindex}
The raw indices are $R_m^{ind} = P_m^{ind}\,C_m$ and $R_m^{col} = P_m^{col}\,f_{frag}$, with $s_0$ set to one since it cancels in the normalisation, each normalised against a fixed reference mission (a typical operational small satellite at \SI{700}{km} sun-synchronous orbit with compliant disposal; its full parameter set is in Table~\ref{tab:inputs}), evaluated in the same scenario,
\begin{equation}
\tilde{R}_m^{\,s} = \frac{R_m^{\,s}}{R_{ref}^{\,s}}, \qquad s\in\{ind,col\}.
\end{equation}
A fixed external reference makes the normalisation stable and $\tilde{R}$ interpretable as a multiple of a typical mission. The continuous index is mapped to five ECSS-compatible classes through logarithmic thresholds,
\begin{equation}
\mathrm{class}(\tilde{R}) = 1 + \sum_{k=1}^{4}\mathbb{1}\!\left[\tilde{R}\ge 10^{\,k-3}\right],
\end{equation}
where $\mathbb{1}[\cdot]$ is the indicator function, equal to one when its argument holds and to zero otherwise, so the five classes span one decade each: class~1 (Very Low, $\tilde{R}<10^{-2}$), class~2 (Low, $10^{-2}\le\tilde{R}<10^{-1}$), class~3 (Medium, $10^{-1}\le\tilde{R}<1$), class~4 (High, $1\le\tilde{R}<10$) and class~5 (Very High, $\tilde{R}\ge 10$). The labels mirror the five-level magnitude scale of the ECSS risk matrix \citep{ECSS_M_ST_80C}. The simultaneous use of the continuous index and the discrete class bridges the categorical and continuous traditions: project teams read the class, sustainability analysts aggregate the index.

\subsection{Implementation}
\begin{figure*}[!t]
\centering
\begin{tikzpicture}[
  >={Latex[length=2mm,width=1.5mm]},
  node distance=8mm,
  every node/.append style={font=\footnotesize},
  io/.style   ={rectangle, rounded corners=2pt, draw=black!70, fill=blue!8,
                align=center, inner xsep=4pt, inner ysep=3pt, minimum width=52mm},
  proc/.style ={rectangle, draw=black!70, fill=orange!12,
                align=center, inner xsep=4pt, inner ysep=3pt, minimum width=52mm,
                font=\footnotesize\ttfamily},
  flux/.style ={rectangle, draw=black!70, fill=orange!12,
                align=center, inner xsep=3pt, inner ysep=3pt, minimum width=36mm,
                font=\footnotesize\ttfamily},
  result/.style={rectangle, rounded corners=2pt, draw=black!70, fill=green!12,
                align=center, inner xsep=4pt, inner ysep=3pt, minimum width=68mm}
]
\node[io]                                       (cat) {Celestrak GP catalogue \\ (active satellites + tracked debris)};
\node[proc, below=of cat]                       (db)  {databasecreator\_real};
\node[io,   below=of db]                        (pop) {Resident population $(h_p,\,\sigma,\,A,\ldots)$};
\node[flux, below left =9mm and 1mm of pop]    (ind) {individual\_ \\ probability\_flux};
\node[flux, below right=9mm and 1mm of pop]    (col) {collective\_ \\ probability};
\node[proc, below=24mm of pop]                  (ri)  {risk\_index};
\node[result, below=of ri]                      (cls) {Continuous $\tilde{R}$ \;+\; discrete class 1--5 (VL--VH)};

\draw[->] (cat) -- (db);
\draw[->] (db)  -- (pop);
\draw[->] (pop) -- (ind);
\draw[->] (pop) -- (col);
\draw[->] (ind) -- (ri);
\draw[->] (col) -- (ri);
\draw[->] (ri)  -- (cls);
\end{tikzpicture}
\caption{Toolchain architecture of the open MATLAB implementation. The Celestrak General Perturbations catalogue is processed into a resident population; the two flux modules share the density formulation~\eqref{eq:density} and differ in the population subset and time horizon (all catalogued objects over $T_{op}$ for the individual scenario; operational objects over $T_{tot}$ with the fragmentation weight $f_{frag}$ for the collective one); the risk-index module normalises against the fixed reference mission and produces both the continuous index $\tilde{R}$ and the five-class label.}
\label{fig:toolchain}
\end{figure*}
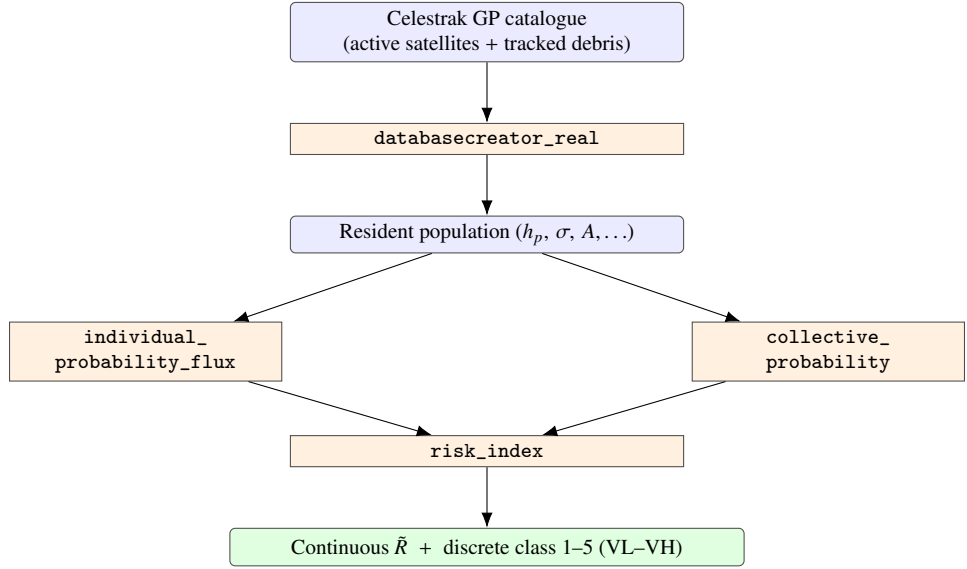

The algorithm is an open MATLAB toolchain. The \texttt{databasecreator\_\allowbreak real} module builds the resident population from the real Celestrak General Perturbations (GP) catalogue (active satellites and tracked debris), deriving each object's perigee altitude from its mean motion and eccentricity. The \texttt{individual\_\allowbreak probability\_\allowbreak flux} and \texttt{collective\_\allowbreak probability} modules implement the two flux formulations, sharing the density computation and differing in population and exposure time (the fragmentation weight of the severity term is applied inside the collective module for convenience, leaving the product $P_m^{col}S_m^{col}$ of Section~\ref{sec:riskindex} unchanged); the \texttt{risk\_\allowbreak index} module normalises and classifies. The flux approach makes the two scenarios consistent and numerically robust across both dense and sparse bands, unlike discrete conjunction methods. The toolchain is released under the MIT licence and deposited on Zenodo (see Data availability); a full three-case evaluation against more than seventeen thousand catalogued objects runs in seconds. The toolchain architecture is summarised in Figure~\ref{fig:toolchain}.

\section{Case studies}
\label{sec:cases}

\subsection{Three representative cases}

\begin{figure}[ht]
\centering
\includegraphics[width=\columnwidth,keepaspectratio]{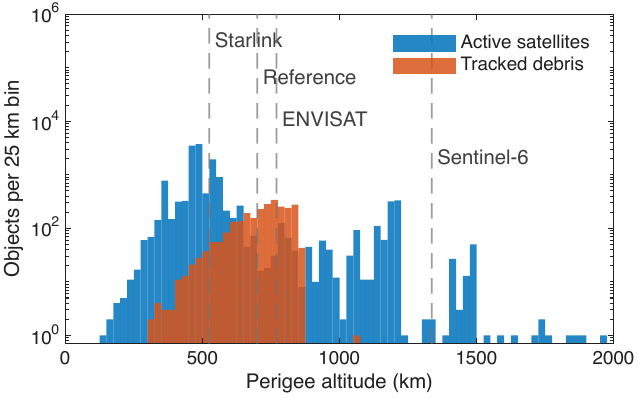}
\caption{Perigee-altitude distribution from the Celestrak catalogue (25~km bins, log count): active satellites (blue) and tracked debris (orange, the four fragmentation clouds). Dashed lines mark the three case studies and the reference mission. The active peak at 525~km and the debris band at 700--850~km drive the case-study results.}
\label{fig:altitude}
\end{figure}

The algorithm is applied, using the real Celestrak catalogue for epoch May 2026 (\num{14727} active LEO satellites and \num{2597} tracked debris objects, the latter comprising the four principal catalogued fragmentation clouds, namely Fengyun-1C, Iridium-33, Cosmos-2251 and Cosmos-1408), to three representative classes of LEO object: a large defunct platform (ENVISAT, \SI{770}{km}), an operational Earth-observation mission (Sentinel-6 Michael Freilich, \SI{1336}{km}) and a constellation element (Starlink V2 Mini, \SI{525}{km}). Orbital states are from public Two-Line Element (TLE) sets \citep{Celestrak2026}; replacement costs from representative commercial valuations \citep{IDA2023}; mission data from \citep{ESA_ENVISAT,ESA_NASA_Sentinel6,FCC_StarlinkV2_2023}. Table~\ref{tab:inputs} lists the input parameters and Table~\ref{tab:results} reports the resulting indices. The perigee-altitude distribution of the catalogue, with the case-study altitudes marked, is shown in Figure~\ref{fig:altitude}.

\begin{table*}[!t]
\centering
\footnotesize
\setlength{\tabcolsep}{4pt}
\caption{Input parameters for the three case studies and the fixed reference mission. $T_{op}$: operational lifetime; $T_{res}$: residual orbital lifetime after disposal; $\sigma$: exposed collision cross-section; $A$: total surface (fragmentation proxy); $C$: replacement cost. Cost figures are first-order estimates of different cost types as available in public sources: ENVISAT $2.3\,\mathrm{B\,USD}$ is the original 2002 build cost \citep{ESA_ENVISAT}; adjusted to 2026 price levels (US consumer-price inflation) it would be roughly $1.7$ times larger, which would raise the ENVISAT individual index by the same factor without changing its class (already saturated at VH) or any conclusion; Sentinel-6 $0.4\,\mathrm{B\,USD}$ is the programme cost \citep{ESA_NASA_Sentinel6}; Starlink V2 Mini $1\,\mathrm{M\,USD}$ is the unit marginal cost class estimated from public statements \citep{FCC_StarlinkV2_2023}; the reference mission cost is the order-of-magnitude class of mid-size operational small satellites \citep{IDA2023}. The normalised indices and the threshold construction of Section~\ref{sec:parametric} are invariant under a common rescaling of all costs; only the ratios $C_m/C_{ref}$ enter the asymmetry.}
\label{tab:inputs}
\begin{tabular}{lcccccc}
\toprule
Object & $h$ (km) & $T_{op}$ (yr) & $T_{res}$ (yr) & $\sigma$ (m$^2$) & $A$ (m$^2$) & $C$ (USD) \\
\midrule
ENVISAT (defunct)            & 770  & 10  & 200 & 100 & 250 & \num{2.3e9} \\
Sentinel-6 (operational)     & 1336 & 5.5 & 25  & 10  & 25  & \num{4.0e8} \\
Starlink V2 (constellation)  & 525  & 5   & 5   & 12  & 30  & \num{1.0e6} \\
Reference (smallsat)         & 700  & 7   & 25  & 5   & 15  & \num{1.0e8} \\
\bottomrule
\end{tabular}
\end{table*}

\begin{table*}[!t]
\centering
\small
\setlength{\tabcolsep}{5pt}
\caption{Risk-index values and discrete classification for the three case studies, computed from the real Celestrak population and normalised against the fixed reference mission. The individual index is the expected-loss risk to the operator (impact probability weighted by replacement value); the collective index is the environmental risk imposed on the operational population, independent of the mission's own value. $R=\tilde{R}^{col}/\tilde{R}^{ind}$ quantifies the operator-versus-environment asymmetry and is computed from the unrounded indices; the apparent ratio from the rounded $\tilde{R}^{ind}/\tilde{R}^{col}$ columns may therefore differ slightly. $^{\ast}$The ENVISAT individual index uses the original build cost as $C_m$ and is therefore retrospective (the platform has been defunct since 2012 and no active replacement investment is in place); for defunct platforms the collective axis is the operationally relevant one for present-day decisions, while the individual axis is reported for completeness. Risk classes: VL~(1) very low, L~(2) low, M~(3) medium, H~(4) high, VH~(5) very high.}
\label{tab:results}
\begin{tabular}{lcccccc}
\toprule
Case study & $h$ (km) & $\tilde{R}^{ind}$ & Class$^{ind}$ & $\tilde{R}^{col}$ & Class$^{col}$ & $R$ \\
\midrule
ENVISAT (defunct$^{\ast}$)  & 770  & 1018 & VH (5) & 903  & VH (5) & 0.9 \\
Sentinel-6 (operational)    & 1336 & 0.02 & L (2)  & 0.05 & L (2)  & 2.1 \\
Starlink V2 (constellation) & 525  & 0.09 & L (2)  & 30   & VH (5) & 317 \\
\bottomrule
\end{tabular}
\end{table*}

\begin{figure}[ht]
\centering
\includegraphics[width=0.88\linewidth]{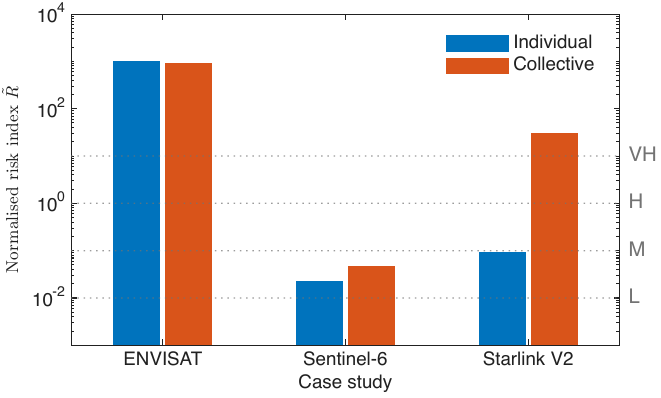}
\caption{Normalised individual and collective risk indices $\tilde{R}$ (logarithmic axis) for the three case studies. Dotted lines mark the class thresholds ($10^{-2}$, $10^{-1}$, $10^{0}$, $10^{1}$); the class entered above each line is indicated on the right. For the Starlink element the collective index exceeds the individual index by more than two orders of magnitude.}
\label{fig:asymmetry}
\end{figure}

The asymmetry between the individual and collective indices is sharpest for the constellation element. For Starlink the individual index is Low ($0.09$) and the collective is Very High ($30$); because both are multiples of the same reference, an individual-only assessment understates the environmental burden by over two orders of magnitude (Figure~\ref{fig:asymmetry}). A cheap satellite, individually low-risk, can dominate the collective degradation of its band when operating in the most congested LEO region. The index is computed per element, with no constellation-level weighting: the value of $30$ refers to a single satellite, and the aggregate burden of a shell populated by thousands of such elements scales linearly with their number, so the per-element asymmetry is a conservative statement of the constellation problem. ENVISAT is Very High on both axes, as expected for a massive defunct platform in a debris-rich band; for a defunct object the individual axis is retrospective, so the collective is the operationally relevant one. Sentinel-6 is Low on both, reflecting a well-designed mission in a sparsely populated regime. The two indices measure different quantities and should not be conflated: a purely economic, operator-centred assessment misses the external cost that large constellations of low-value satellites place on the orbital commons. Operationally, the three cases map onto three uses: ENVISAT as an active debris removal priority consistent with ESA's Zero Debris position \citep{ESA_ZeroDebris_2023}; Sentinel-6 as a low-risk baseline; Starlink as a regulatory case, given that current guidelines \citep{IADC2021} address the lifetime collective contribution of constellation elements only indirectly through end-of-life disposal.

\subsection{Statistical validation across the catalogue}
\label{sec:validation}

The three cases above span the operational LEO range but do not, by themselves, establish that the individual--collective asymmetry is a structural property of the current catalogue and not a feature of the three chosen configurations. To address this, the algorithm is evaluated on a random sample of $N=100$ active LEO satellites drawn from the same May 2026 Celestrak GP snapshot (\num{14727} eligible objects, with the random number generator seeded as \texttt{rng(20260605, 'twister')} for byte-identical reproduction from the public Zenodo deposit). Each sampled satellite inherits its orbital state from the catalogue.

Non-orbital parameters (replacement cost, exposed and total surface, operational and residual lifetime) are drawn from an altitude-stratified prior intended to reflect the typical configuration of the three tiers identifiable in the current LEO population: small constellation buses below~\SI{600}{km} (cost log-normal with median~\SI{5e6}{USD}, $\sigma_{\log}=0.5$), medium Earth-observation and science platforms between 600 and~\SI{1500}{km} (median~\SI{1.5e8}{USD}, $\sigma_{\log}=0.7$), and large specialty platforms above~\SI{1500}{km} (median~\SI{4e8}{USD}, $\sigma_{\log}=0.5$). Cross-section and lifetime priors follow the same tiering. The reference mission of Section~\ref{sec:cases} is reused without modification, so individual and collective indices remain interpretable as multiples of the same baseline.

\begin{figure*}[!t]
\centering
\includegraphics[width=0.97\linewidth]{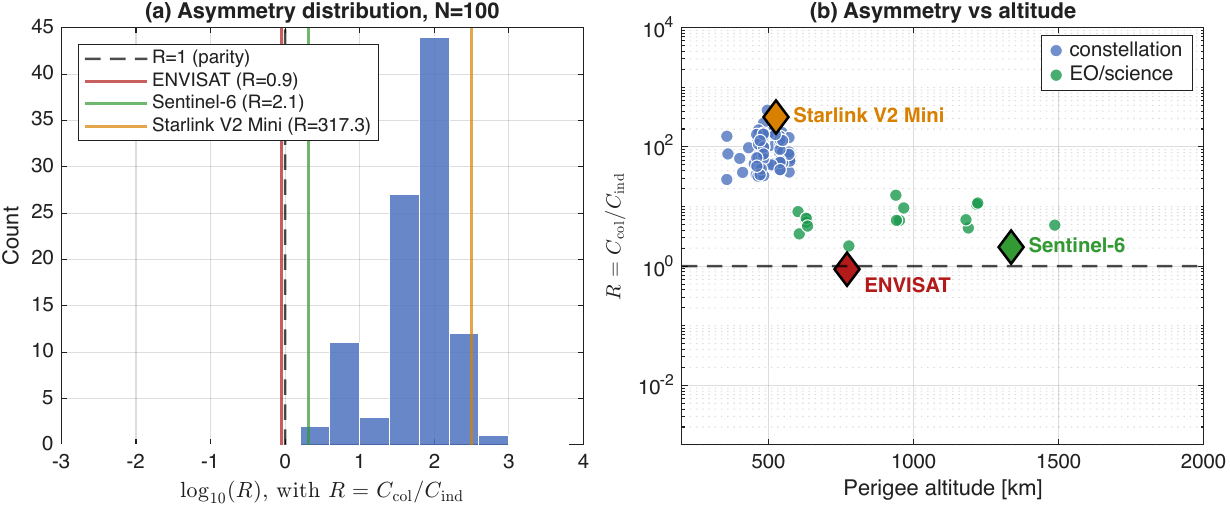}
\caption{Statistical validation of the individual--collective asymmetry on a random sample of $N=100$ active LEO satellites from the May 2026 Celestrak catalogue. (a) Distribution of $\log_{10}(R)$ across the sample. ENVISAT ($R\!\approx\!0.9$, a large defunct platform whose own replacement value lifts $\tilde{R}^{ind}$ toward parity with $\tilde{R}^{col}$) and Sentinel-6 ($R\!\approx\!2.1$, sparse high band) sit in the left tail; Starlink V2~Mini ($R\!\approx\!317$) sits near the $p_{99}$, not as an outlier but as the upper end of a populated cluster. (b) $R$ versus perigee altitude, colour-coded by altitude tier. The constellation tier ($h<\SI{600}{km}$) clusters at $R\!\in\![50,200]$; the EO/science tier clusters one to two orders of magnitude lower. The specialty tier ($h\geq\SI{1500}{km}$) is unrepresented in this random sample, consistent with the current catalogue composition.}
\label{fig:validation}
\end{figure*}

Across the sample (seed \texttt{rng(20260605)}), $R = \tilde{R}^{col}/\tilde{R}^{ind}$ has median $73$, inter-quartile range $[42,126]$, $p_{90}=163$ and $p_{99}=385$; $R>1$ on all $100$ sampled objects, giving a Clopper-Pearson 95\% confidence lower bound $P(R>1)\geq 0.964$ on the underlying population proportion. The sample size is set by the precision the claim needs: with every draw positive, one hundred objects already pin the binomial lower bound above $96\%$; the hundred-seed sweep below, ten thousand sampled objects in all, bounds seed noise; and the uniform draw reproduces the catalogue's own tier composition, which is the population the structural claim is about. A seed sweep across $100$ independent draws of the validation sample (seeds $20260605$ to $20260704$) yields median $R\in[63,95]$ with sweep median $78$ and $p_{05}{-}p_{95}$ interval $[68,91]$, $P(R>1)$ with sweep median $0.99$ and range $[0.95, 1.00]$, and the Spearman rank correlation $\rho_{S}$ between the collective index and the snapshot-ECOB proxy (Section~\ref{sec:ecob-bench}) in $[0.84,0.94]$ with sweep median $0.91$: the headline numbers are stable to within seed-noise. The asymmetry $R\gg 1$ is therefore an algebraic consequence of the homogeneous-severity design choice combined with the empirical cost distribution of LEO satellites: the design choice predicts it analytically (Section~\ref{sec:parametric} eq.~\ref{eq:asymmetry-cost}), and the sample confirms it numerically. The Starlink V2~Mini ratio of $\sim\!317$ (Section~\ref{sec:cases}) sits near the sample $p_{99}$; the upper-tail cluster lies entirely between 465 and~\SI{546}{km}, the constellation band, and is dominated by Starlink-class objects (Table~\ref{tab:top10}, Figure~\ref{fig:validation}). The dominance of the constellation tier (\num{84} of \num{100} objects) reflects the current catalogue composition; the EO/science tier alone ($n=16$, median $R=6$, $p_{90}=11$) shows a smaller but non-zero asymmetry, the specialty tier ($h\ge\SI{1500}{km}$) is unrepresented in the May~2026 sample, so the strong-asymmetry claim generalises empirically only within the constellation and EO/science bands.

The asymmetry is robust to both the cost prior and the snapshot composition. Scaling every sampled cost by a factor $\xi$ rescales the median asymmetry as $R(\xi)=R(1)/\xi$ (at $\xi=0.5$, $R=147$; at $\xi=1.5$, $R=49$), preserving the qualitative claim, while the replacement-cost threshold $C^{*}(h)$, the cost below which the collective--individual asymmetry exceeds two orders of magnitude (introduced in Section~\ref{sec:parametric}), is by design prior-independent. Repeating the analysis on a synthetic past-2025 snapshot, built from the live June~2026 catalogue by removing all 2026-launched objects (about $1900$ predominantly Starlink V2 Mini elements, leaving \num{12962} active LEO) preserves median $R=74.1$ (vs $73.4$), $p_{99}=310$ (vs $393$), $P(R>1)=99\%$ (vs $100\%$), and marginally strengthens the proxy benchmark to $\rho_{S}=0.92$ (vs $0.91$). The 2026 launches amplify but do not create the asymmetry the paper describes.

\begin{table*}[!t]
\centering
\footnotesize
\setlength{\tabcolsep}{6pt}
\caption{Top-10 individual--collective asymmetry $R = \tilde{R}^{col}/\tilde{R}^{ind}$ across the $N=100$ random sample. All ten upper-tail objects lie within~\SI{82}{km} of one another in the constellation band (465 to~\SI{546}{km}); replacement costs are sampled from the constellation-tier prior (median~\SI{5e6}{USD}). The Starlink V2~Mini value of Section~\ref{sec:cases} ($R\!\approx\!317$, individual class L, collective class VH) lies at the same operating point. Object names are the Celestrak \texttt{OBJECT\_NAME} field as of 29 May 2026.}
\label{tab:top10}
\begin{tabular}{lrrr}
\toprule
Object & Altitude (km) & Cost (MUSD) & $R$ \\
\midrule
2024-205L                & 495 & 1.34 & 413 \\
STARLINK-3892            & 539 & 1.41 & 356 \\
STARLINK-35569           & 482 & 2.39 & 250 \\
STARLINK-36792           & 465 & 2.04 & 193 \\
TK-2                     & 546 & 2.91 & 175 \\
SPACEMOBILE-006          & 502 & 2.56 & 171 \\
STARLINK-32329           & 482 & 3.05 & 171 \\
STARLINK-32162           & 475 & 3.16 & 171 \\
TRANSPORTER-15 OBJECT DM & 505 & 2.76 & 164 \\
STARLINK-6141            & 482 & 3.20 & 164 \\
\bottomrule
\end{tabular}
\end{table*}

\subsection{Parametric extension and cost-threshold regulation}
\label{sec:parametric}

\begin{figure}[ht]
\centering
\includegraphics[width=0.92\linewidth]{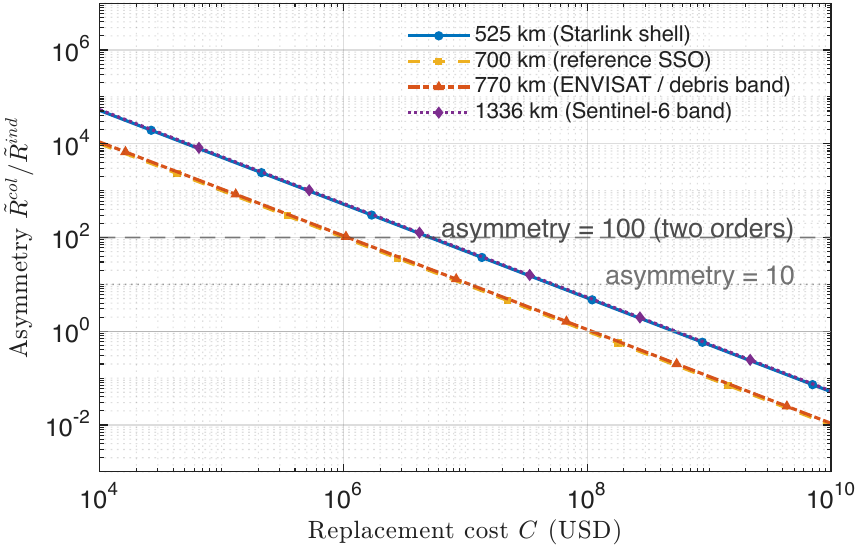}
\caption{Normalised asymmetry $\tilde{R}^{col}/\tilde{R}^{ind}$ as a function of replacement cost $C$, for four altitudes, with satellite geometry and lifetime held at the reference values. The relation is linear with slope $-1$ in log--log; the 525 and 1336~km curves overlap visually, as do the 700 and 770~km curves, reflecting the U-shape of $K(h)$ discussed in the text. Dashed and dotted horizontal lines mark the asymmetry levels of 100 (two orders of magnitude) and 10.}
\label{fig:parametric}
\end{figure}

The three cases above cover specific operating points but leave open whether the individual--collective asymmetry is structural or peculiar to those configurations. To isolate the cost dependence, the satellite geometry and lifetime are held at the reference values ($\sigma=5\,\mathrm{m^2}$, $A=15\,\mathrm{m^2}$, $T_{op}=7\,\mathrm{yr}$, $T_{res}=25\,\mathrm{yr}$) and only the replacement cost $C$ is varied. Under this constraint the normalised asymmetry depends on cost exactly as
\begin{equation}
\frac{\tilde{R}^{col}}{\tilde{R}^{ind}}(h,C) = \frac{K(h)}{C}, \qquad
K(h) = C_{ref}\cdot\frac{P_{col}(h)}{P_{col,ref}}\cdot\frac{P_{ind,ref}}{P_{ind}(h)},
\label{eq:asymmetry-cost}
\end{equation}
a straight line of slope $-1$ in the log--log plane whose vertical intercept $K(h)$ depends only on the altitude through the ratio between the collective and the individual collision rates. Figure~\ref{fig:parametric} shows the relation for four representative altitudes; Table~\ref{tab:thresholds} reports, for each altitude, the normalised collective index of a reference-sized satellite (independent of cost) and the cost threshold $C^{*}$ below which the asymmetry exceeds two orders of magnitude.

\begin{table*}[!t]
\centering
\small
\setlength{\tabcolsep}{6pt}
\caption{Altitude-resolved analysis of the asymmetry. $\tilde{R}^{col}_{\,ref}$ is the normalised collective index of a reference-sized satellite at altitude $h$ (independent of cost); the class column maps it onto the ECSS-compatible scheme; $C^{*}$ is the replacement-cost threshold below which the asymmetry $\tilde{R}^{col}/\tilde{R}^{ind}$ exceeds two orders of magnitude.}
\label{tab:thresholds}
\begin{tabular}{lccc}
\toprule
Altitude (km) & $\tilde{R}^{col}_{\,ref}$ & Class & $C^{*}$ (USD) \\
\midrule
525  & 28.2  & VH (5) & \num{5.1e6} \\
700  & 1.00  & H (4)  & \num{1.0e6} \\
770  & 1.69  & H (4)  & \num{1.1e6} \\
1336 & 0.019 & L (2)  & \num{5.3e6} \\
\bottomrule
\end{tabular}
\end{table*}

Two features of \eqref{eq:asymmetry-cost} matter for regulation. First, $K(h)$ is U-shaped in altitude: it peaks in the most congested band (525~km, Starlink shell) and in the sparsest one (1336~km, Sentinel-6 band), and is lowest at the intermediate, debris-rich bands (700--770~km) where the high all-object density partially compensates the operational density. Second, the asymmetry alone is not the regulatory quantity of interest: at 1336~km the reference-sized collective index already sits at class L (the ECSS-compatible scheme of Section~\ref{sec:riskindex}), so the asymmetry is structural but operationally irrelevant. The regulatorily significant zone is the conjunction of two conditions: a reference-sized collective index already at class H or VH (altitudes 525, 700, 770~km) \emph{and} a replacement cost below $C^{*}$. Within this zone the operator-centred assessment classifies the mission as Low or below, while the collective assessment classifies it as High or Very High.

At the Starlink shell, any satellite below approximately five million USD in replacement value is therefore structurally individual-Low and collective-Very High, given the reference mission of Section~\ref{sec:cases} ($C_{ref}=\SI{e8}{USD}$). $C^{*}(h)$ scales linearly with $C_{ref}$: equation~\eqref{eq:asymmetry-cost} gives $K(h)\propto C_{ref}$, so all thresholds rescale by the same factor under a different reference. With $C_{ref}=\SI{25}{}$, $\SI{50}{}$, $\SI{100}{}$, $\SI{200}{MUSD}$, the 525~km threshold becomes $1.3$, $2.6$, $5.1$, $\SI{10.2}{MUSD}$; the 700~km becomes $0.25$, $0.50$, $1.0$, $\SI{2.0}{MUSD}$. The relative statement is invariant: a mission with replacement cost below $5\%$ of the reference is structurally individual-Low and collective-VH at the Starlink shell, irrespective of which reference is chosen. The Starlink V2~Mini placement is therefore robust to the reference choice within the typical small-satellite range ($C_{ref}\in[25,200]\,\mathrm{MUSD}$): its $\sim\!\SI{e6}{USD}$ cost sits below the threshold at the Starlink shell for $C_{ref}>\SI{20}{MUSD}$. Current debris-mitigation guidelines \citep{IADC2021} constrain risk at the individual level and disposal at end of life, but do not address the collective lifetime-integrated contribution; the thresholds in Table~\ref{tab:thresholds} quantify the regime they leave unconstrained. The companion ECOB index \citep{Letizia2016} captures the same physical effect through long-term debris-evolution modelling; the construction here makes the threshold accessible without long-term simulation. The boundary aligns with the economic argument for altitude-graduated orbital-use fees \citep{Rao2020} but represents a categorical threshold, not a continuous Pigouvian price: $C^{*}(h)$ identifies the missions that a flat-fee or per-launch policy would systematically under-price, while a graduated fee following Rao et al.~\citep{Rao2020} would internalise the externality continuously and would not require $C^{*}$.

A consequence of equation~\eqref{eq:asymmetry-cost} concerns the small-satellite range, with one caveat: the cross-section $\sigma_m$ cancels in the ratio, while the fragmentation weight $f_{frag}$ enters the collective index only. A kilogram-scale spacecraft therefore does not inherit the constellation asymmetry unchanged: the smaller fragmentation footprint attenuates it, the lower replacement cost amplifies it. Re-running the algorithm at the Starlink shell with representative educational geometries quantifies the balance. A university-class 3U CubeSat ($f_{frag}\approx 0.10$, $T_{op}=\SI{2}{yr}$, $T_{res}=\SI{6}{yr}$, replacement cost \SI{1}{MUSD}) yields $R\approx 43$ with a geometry-specific threshold $C^{*}\approx\SI{0.43}{MUSD}$; a 2P PocketCube at \SI{50}{kUSD} yields $R\approx 460$, above the Starlink V2~Mini value, because the cost amplification outweighs the footprint attenuation. The absolute collective index of such objects remains in class VL, so band degradation is driven by constellation hardware, not by educational payloads; the asymmetry diagnosis, however, transfers unchanged. The cheaper the spacecraft, the wider the gap between the risk its operator bears and the risk it imposes: educational missions in congested bands are the limiting case of the regime that operator-centred frameworks underrate.

\section{Discussion}
\label{sec:discussion}

The classifications agree, qualitatively, with the published debris-risk literature. ENVISAT is the upper baseline of the ECOB family of indices~\citep{Letizia2016}; the present construction places it at class VH on both axes, in agreement with that ranking. Starlink V2~Mini at \SI{525}{km} classifies as Very High on the collective axis, consistent with the empirically documented surge of collision-avoidance manoeuvres at this shell~\citep{ESA_SpaceEnv_2024}; Sentinel-6 at \SI{1336}{km} classifies as Low on both axes. The statistical evaluation of Section~\ref{sec:validation} shows that this ordering generalises within the constellation and EO/science tiers of the catalogue; Section~\ref{sec:ecob-bench} reports the proxy sensitivity check and the three-point literature anchor against published ECOB values. The continuous index supports comparative ranking without the rounding of categorical classification; the collective index provides a basis for prioritising active debris removal candidates, with ENVISAT consistent with the established remediation consensus \citep{LiouJohnson2008}; and the individual--collective asymmetry differentiates mission-level mitigation from constellation-level sustainability requirements.

Current debris-mitigation guidelines constrain risk at the per-mission level (collision-probability thresholds, $25$-year disposal, more recently $5$-year for new constellations) but do not constrain the lifetime-integrated collective contribution of a constellation element. A supplementary, altitude-dependent requirement consistent with the framework of this paper is therefore proposed: \emph{a satellite whose normalised individual index $\tilde{R}^{ind}$ is below the L/M class boundary and whose normalised collective index $\tilde{R}^{col}$ is at or above the H/VH class boundary triggers a constellation-level review at the licensing stage}. Operationally, this translates into a clause that could be added to IADC guideline~5.3, ECSS-U-AS-10C section on disposal, or to FCC orbital-debris mitigation rules for non-geostationary licences (\S25.114 of the FCC commercial licensing pathway): for any element of a constellation operating below the altitude-resolved cost threshold $C^{*}(h)$ of Table~\ref{tab:thresholds}, the operator demonstrates either a per-element mitigation measure (active debris removal commitment, end-of-life thrust budget, controlled deorbit) or accepts a constellation-aggregated cap on the integral $\sum_i \tilde{R}_i^{col}$. The threshold itself is computable from a catalogue snapshot in seconds and updates automatically as the catalogue evolves, addressing the practical difficulty of recalibrating policy as the operational population grows.

Two incentive objections bound the design space of such a clause. First, the trigger can be gamed: over-declaring the replacement cost raises the individual index and can lift an element above the L/M boundary, out of the trigger's reach. The exposure is limited by anchoring $C_m$ to the insured value where the asset is insured (over-declaration then costs premiums) and, for the self-insured fleets typical of large constellations, to the unit costs disclosed in licensing and financial filings (over-declaration then contradicts the operator's own filings); it vanishes altogether under the constellation-aggregated alternative, since the collective index is cost-independent by construction. An orbital-use fee \citep{Rao2020} avoids declared quantities altogether, at the price of requiring a fee authority that the licensing trigger does not need. Second, any licensing threshold can act as a barrier to entry if incumbents shape it \citep{Adilov2015}; here the trigger binds only low-unit-cost elements of large constellations in already-congested bands, so a small entrant with few satellites faces either no trigger or a small aggregate cap, while the fleets the review addresses are precisely the incumbents' own. The distributional design of the associated requirements, fees or caps, is policy work beyond the scope of this paper; the index contributes the observable such a design would act on.

\subsection{Internal sensitivity and external literature anchor}
\label{sec:ecob-bench}

The construction is exposed to two complementary checks. The first is an \emph{internal sensitivity check}: a snapshot-only proxy of the ECOB family is implemented, sharing the same kinetic-flux density of $P_m^{col}$ but with three ECOB-aligned modifications: (i)~a mass-based fragmentation weight $f_{frag}^{eco}=(M_m/M_{ref})^{0.75}$ following the NASA Standard Breakup Model fragment-count scaling adopted in ECOB \citep{Letizia2016,Johnson2001}, in place of the area-based exponent~$0.5$ used in $S_m^{col}$; (ii)~a fixed long-term horizon $T_{eco}=\SI{200}{yr}$ in place of $T_{op}+T_{res}$; (iii)~snapshot density held at the catalogue value, no evolution. The proxy is implemented as the module \texttt{ecob\_proxy.m} in the public Zenodo deposit. The two indices therefore share their density backbone and differ only in fragmentation weight and time horizon, so the agreement between them measures the sensitivity of the ranking to those two choices, not external validation. External validation comes from the \emph{literature anchor} reported below. Neither check substitutes the full ECOB toolchain, which requires long-term Monte Carlo evolution \citep{Letizia2016,Letizia2017} and is the subject of a forthcoming multi-domain framework by the author.

The three case studies illustrate the effect of the two ECOB-aligned modifications. Table~\ref{tab:ecob-cases} reports $\tilde{R}^{col}$ and $\tilde{R}^{eco}$ side by side. Both indices place ENVISAT in class VH, Starlink V2~Mini in the upper range and Sentinel-6 in class L, reproducing the ordering ENVISAT $>$ Starlink V2~Mini $>$ Sentinel-6 documented in the ECOB literature with ENVISAT as the upper baseline \citep{Letizia2016}. The magnitudes differ in informative ways: the proxy assigns ENVISAT a lower value than $P_m^{col}$ ($\tilde{R}^{eco}/\tilde{R}^{col}=0.30$) because mass-based weighting under-emphasises very large platforms relative to area-based weighting, and assigns Starlink V2~Mini a higher value ($3.22$) because the 200-year horizon amplifies dense-band contributions. Sentinel-6 is essentially unchanged ($1.56$).

The same comparison answers a formulation question directly: whether the individual--collective asymmetry of Section~\ref{sec:cases} is an artefact of the area-based severity choice. Recomputing the asymmetry with the mass-based weight, $R^{eco}=\tilde{R}^{eco}/\tilde{R}^{ind}$, preserves the ordering and widens the constellation gap: ENVISAT moves from $0.9$ to $0.27$, Sentinel-6 from $2.1$ to $3.3$, and Starlink V2~Mini from $317$ to over a thousand, since the mass-based exponent weights an 800-kg element more heavily than its area does. The five-class assignments of Table~\ref{tab:results} are unchanged under the swap, as they are under the exponent sweep $\alpha\in[0.3,0.75]$ of Section~\ref{sec:limitations} and under the uniform cost rescaling of Section~\ref{sec:validation}. The asymmetry, and the regulatory reading built on it, is a property of pairing an asset-weighted individual index with a value-independent collective one, not of the particular severity proxy chosen; Section~\ref{sec:limitations} states the converse, that an asset-weighted collective index would erase it by construction.

\begin{table*}[!t]
\centering
\footnotesize
\setlength{\tabcolsep}{6pt}
\caption{Snapshot-ECOB proxy benchmark on the three case studies. $\tilde{R}^{col}$ is the normalised collective index of this paper (area-based fragmentation weight, operational plus residual horizon); $\tilde{R}^{eco}$ is the snapshot-ECOB proxy defined above (mass-based NASA-Standard-Breakup fragmentation weight, 200-year horizon, same catalogue snapshot). Both indices are normalised against the reference mission ($M_{ref}=\SI{500}{kg}$, \SI{700}{km} sun-synchronous). Object masses are from public sources: ENVISAT \SI{8211}{kg} (ESA fact sheet), Sentinel-6 Michael Freilich \SI{1192}{kg} (NASA/JPL fact sheet), Starlink V2~Mini \SI{800}{kg} (SpaceX engineering value).}
\label{tab:ecob-cases}
\begin{tabular}{lrrrr}
\toprule
Case study & $M$ (kg) & $\tilde{R}^{col}$ & $\tilde{R}^{eco}$ & ratio $\tilde{R}^{eco}/\tilde{R}^{col}$ \\
\midrule
ENVISAT          & 8211 & 903.19 & 275.01 & 0.30 \\
Sentinel-6       & 1192 &   0.05 &   0.07 & 1.56 \\
Starlink V2 Mini &  800 &  29.95 &  96.39 & 3.22 \\
\bottomrule
\end{tabular}
\end{table*}

\begin{figure}[ht]
\centering
\includegraphics[width=0.88\linewidth]{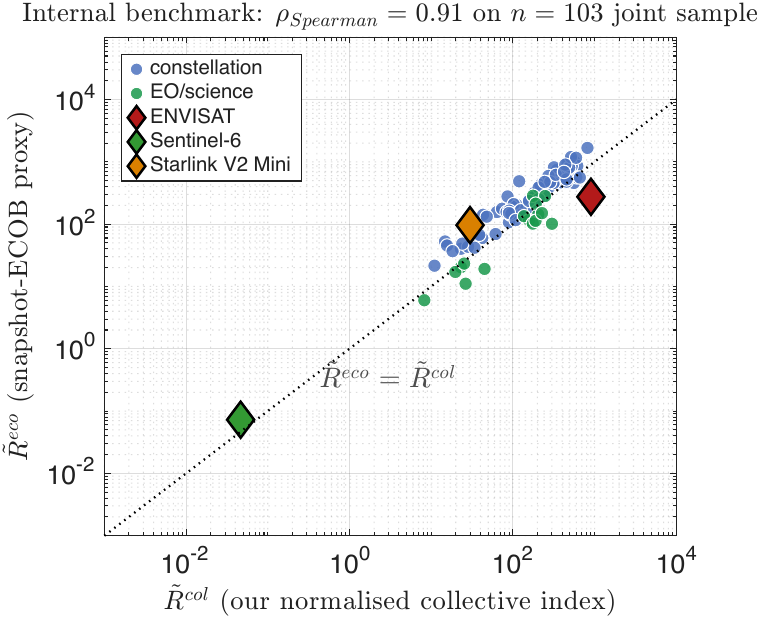}
\caption{Snapshot-ECOB proxy benchmark on the joint sample of three case studies (diamonds) and $N=100$ random LEO satellites (circles, coloured by altitude tier). The dotted diagonal marks the identity $\tilde{R}^{eco}=\tilde{R}^{col}$; the Spearman rank correlation over the joint sample is $\rho_{S}=0.91$. Both indices are computed on the same May 2026 Celestrak snapshot. The systematic offset of the case studies along the diagonal isolates the simplifications that distinguish $P_m^{col}$ from $\tilde{R}^{eco}$: ENVISAT sits below identity because the mass-based fragmentation weight under-emphasises very large platforms; Starlink V2~Mini sits above identity because the 200-year horizon amplifies dense-band contributions; Sentinel-6 sits on identity because its sparse-band operating point is unaffected by either choice.}
\label{fig:ecob-bench}
\end{figure}

Extending the comparison to the $N=100$ random sample of Section~\ref{sec:validation} (masses drawn from an altitude-stratified log-normal prior: constellation \SI{500}{kg} median, EO/science \SI{1500}{kg}, specialty \SI{3000}{kg}) gives a Spearman rank correlation $\rho_{S}=0.91$ between $\tilde{R}^{col}$ and $\tilde{R}^{eco}$ on the joint sample $n=103$ (Figure~\ref{fig:ecob-bench}). Per-tier: $\rho_{S}=0.93$ constellation ($n=84$), $\rho_{S}=0.75$ EO/science ($n=16$, Fisher 95\% CI $[0.40,0.91]$, wider due to mass and surface dispersion). Across a sweep of $100$ independent rng draws of the validation sample, $\rho_{S}$ lies in $[0.84, 0.94]$ (median $0.91$). Because the two indices share their density backbone, $\rho_{S}$ measures the sensitivity of the ranking to the choice of fragmentation weight and time horizon, not external benchmark consistency; that consistency is provided by Table~\ref{tab:literature-anchor} below.

The ratio $\tilde{R}^{eco}/\tilde{R}^{col}$ diverges in two predictable directions: for massive objects in moderately populated bands ($<1$, ENVISAT regime) the area-based exponent overweights the platform; for low-mass objects in dense bands ($>1$, Starlink regime) the 200-year horizon under-weights it. The ratio would constitute, in the companion multi-domain framework, the first-order calibration of $P_m^{col}$ against the full ECOB.

The proxy is benchmarked against published ECOB values. The study presented by Letizia et al.\ at the 7th European Conference on Space Debris (SDC7)~\citep{Letizia2017} reports debris-risk indices normalised against ENVISAT at the 2016 reference epoch for two well-known EO platforms, MetOp-A and Sentinel-2, both at progressively smaller mass in the ENVISAT altitude band (786 to~\SI{827}{km}). Recomputing $\tilde{R}^{col}$ and $\tilde{R}^{eco}$ on those missions with public engineering parameters and re-normalising against ENVISAT yields the ratios in Table~\ref{tab:literature-anchor}. The proxy $\tilde{R}^{eco}$ matches the published ECOB ratios to within a factor of $\sim\!2$ on both anchors (MetOp-A $0.144$ vs $0.180$; Sentinel-2 $0.025$ vs $0.030$). The area-based index $\tilde{R}^{col}$ ranks identically but is uniformly displaced low: it disagrees by a factor of $\sim\!8$ on MetOp-A and $\sim\!6$ on Sentinel-2, attributable to its area-based exponent and the longer residual-time treatment inherited by ENVISAT in the present framework (\SI{200}{yr} natural decay versus the disposed-mission horizon adopted in the published normalisation). The divergence is a feature of the design choice and is reabsorbed once $\tilde{R}^{eco}$ is used. The literature-anchor sample comprises only three points and is descriptive, not statistical; the qualitative agreement is reported without computing a coefficient.

\begin{table*}[!t]
\centering
\footnotesize
\setlength{\tabcolsep}{6pt}
\caption{Literature-anchor benchmark against published ECOB values \citep{Letizia2017}: published values normalised against ENVISAT at the 2016 epoch (their Table~4); present indices recomputed on the May~2026 snapshot and re-normalised against ENVISAT. Sentinel-2 published value is inferred from the 100-year reference mission profile of \citep{Letizia2017}. Three anchors, descriptive: $\tilde{R}^{eco}$ matches within $\sim\!2\times$ on both non-baseline anchors; $\tilde{R}^{col}$ ranks identically but disagrees by $\sim\!8\times$ on MetOp-A and $\sim\!6\times$ on Sentinel-2, reflecting its area-based weight and the longer residual-time horizon adopted for ENVISAT.}
\label{tab:literature-anchor}
\begin{tabular}{lrrrrr}
\toprule
Mission & $h$ (km) & $M$ (kg) & $\tilde{R}^{col}/\tilde{R}^{col}_\mathrm{ENV}$ & $\tilde{R}^{eco}/\tilde{R}^{eco}_\mathrm{ENV}$ & ECOB$/$ECOB$_\mathrm{ENV}$ \citep{Letizia2017} \\
\midrule
ENVISAT      & 770 & 8211 & 1.000 & 1.000 & 1.000 \\
MetOp-A      & 827 & 4085 & 0.023 & 0.144 & 0.180 \\
Sentinel-2   & 786 & 1145 & 0.005 & 0.025 & 0.030 \\
\bottomrule
\end{tabular}
\end{table*}

\subsection{Limitations}
\label{sec:limitations}
First, the kinetic-flux formulation assumes objects uniformly distributed within the altitude shell and a single mean relative velocity, and captures the average flux through a band but not the fine geometric structure of individual encounters, for which precise conjunction analysis is required \citep{AkellaAlfriend2000,LemmensKrag2014}: the present index is a band-level screening tool, not a conjunction-warning system, and its case studies are demonstrations on public TLE data; operational deployment would ingest operator ephemerides where available. The formulation likewise does not model collision-avoidance capability: operational satellites manoeuvre against predicted conjunctions with tracked objects, so the raw flux overstates the realised impact rate for manoeuvrable spacecraft. Since collision avoidance is near-universal among active operators, the overstatement enters as a roughly uniform scale factor largely absorbed by the normalisation; more fundamentally, the index quantifies the structural load a mission places on its orbital environment, not the effectiveness of the operational mitigations deployed against it, and avoidance is in any case unavailable against the untracked sub-catalogue population. The regulatory reading of Section~\ref{sec:discussion} prices this structural, unmitigated load deliberately: avoidance capability protects the operator's own asset and lasts only while the spacecraft functions, while the load the index measures persists through failure states and past end of life, precisely the states that generate debris; a burden contingent on the residual post-mitigation risk would credit the part of the risk the operator already internalises and exempt the part no operator can mitigate. The index reflects the tracked catalogue and the four principal fragmentation clouds; sub-catalogue fragments are absent, so absolute indices are lower bounds while the relative classification is preserved against the same reference population (Section~\ref{sec:flux}).

\begin{figure}[ht]
\centering
\includegraphics[width=0.85\linewidth]{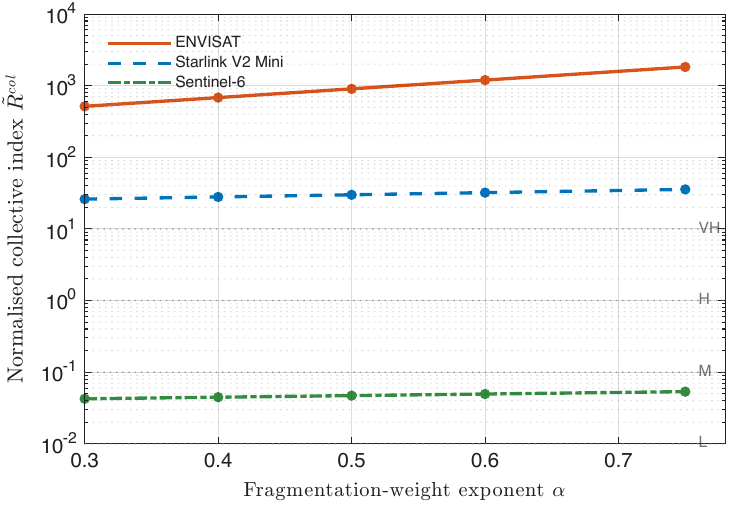}
\caption{Sensitivity of the normalised collective index $\tilde{R}^{col}$ of the three case studies to the fragmentation-weight exponent $\alpha$ in $f_{frag}=(A_m/A_{ref})^{\alpha}$. Markers are the discrete values used in the text ($\alpha\in\{0.30,0.40,0.50,0.60,0.75\}$); the dotted horizontal lines indicate the class thresholds. The discrete five-class classification (ENVISAT VH, Starlink V2 Mini VH, Sentinel-6 L) is invariant across the swept range.}
\label{fig:alpha-sensitivity}
\end{figure}

Second, the homogeneous collective severity is a design choice, not an oversight. An asset-weighted alternative would shift the index toward an operator-portfolio view and erase the asymmetry the construction is designed to expose; the homogeneous choice corresponds to a regulatory view in which environment-level harm depends on how many operational missions $m$ threatens and how large the fragment cloud it produces, not on whether the threatened missions are cheap or expensive. The magnitude of the asymmetry reported in Section~\ref{sec:parametric} follows by construction from this choice: it converts the implicit operator-versus-environment trade-off into an explicit, numerical one. The only free parameter beyond the reference is the exponent $\alpha$ in $f_{frag}=(A_m/A_{ref})^{\alpha}$, fixed at $0.5$; the thresholds $C^{*}(h)$ are $\alpha$-independent by construction (reference geometry yields $f_{frag}=1$), and re-running for $\alpha\in[0.3,0.75]$ yields ENVISAT $\tilde{R}^{col}\in[515,1825]$, Sentinel-6 $\in[0.042,0.053]$ and Starlink V2 $\in[26.1,35.6]$, leaving the five-class classification of Table~\ref{tab:results} invariant (Figure~\ref{fig:alpha-sensitivity}).

Third, the algorithm is a screening layer for prospective classification on public data, not a substitute for operational conjunction analysis or for long-term Monte Carlo evolution \citep{Servadio2024,Simha2025}. The continuous index is compatible by construction with the orbital-sustainability indices reviewed above and can be incorporated as the orbital component of a composite indicator aggregating terrestrial, atmospheric and orbital impacts. The Space Sustainability Rating \citep{Rathnasabapathy2025} aggregates multiple orbital-sustainability inputs into a single mission rating; the present index is intended as a physical screening layer upstream of such ratings. A forthcoming multi-domain framework by the author adopts the risk-based rationale of this work for its orbital component, with an ECOB-class index as the high-fidelity component and the present algorithm as the rapid screening proxy.

\section{Conclusions}

This work has presented a simplified engineering algorithm for the assessment and classification of collision risk in LEO, grounded in the kinetic theory of the debris flux and producing a continuous index for both individual and collective scenarios while retaining compatibility with the categorical ECSS classification. Applied to three representative cases on the May~2026 Celestrak catalogue, it found an individual--collective asymmetry of over two orders of magnitude for the constellation element, showing how operator-centred assessment understates the collective sustainability cost of large constellations. A statistical evaluation on a random sample of \num{100} catalogued LEO satellites confirmed the asymmetry as structural (every sampled object yields collective above individual, sweep-median ratio seventy-eight, the upper decile clustered in the constellation band); a snapshot proxy of the ECOB family benchmarked the construction at Spearman $\rho{=}0.91$ on a joint sample of $103$ objects and matched published ECOB values for MetOp-A and Sentinel-2 to within a factor of two on a literature anchor; the asymmetry is invariant under the swap to a mass-based severity weight, and both checks are stable on a past-2025 synthetic snapshot. A parametric extension gives explicit replacement-cost thresholds per altitude band, naming the regime in which current debris-mitigation guidelines structurally fail to constrain low-value satellites in congested orbits. The algorithm is released as an open toolchain, intended as the orbital input of a multi-domain Space Sustainability Composite Indicator in preparation by the author. Future work will extend the analysis to constellation and fleet scale, incorporate sub-catalogue debris, and integrate with the operational compliance assessment of the Zero Debris regulatory pathway.

\section*{Funding}
The author's post-doctoral position at CISAS ``G.~Colombo'', University of Padova, is funded by Fondazione Cariverona within the Orbis Viridis project (R\&S 2025 call). No other specific funding was received for this work.

\section*{Data availability}
The MATLAB toolchain and the input data (Celestrak GP catalogue snapshots) are openly available on Zenodo at \url{https://doi.org/10.5281/zenodo.20625216} (release v2.0.1, MIT licence), and on GitHub at \url{https://github.com/federicotoson-uni/suslifepath-paper0}. The release reproduces all numerical results, figures and tables of this paper from a fixed random-number seed.

\section*{Declaration of generative AI and AI-assisted technologies in the writing process}
During the preparation of this work the author used Claude (Anthropic) to support language refinement and code testing. After using this tool, the author reviewed and edited the content and takes full responsibility for the content of the publication.

\bibliographystyle{elsarticle-num}
\bibliography{07_References}

\end{document}